\documentclass[12pt]{iopart}
\usepackage{makecell}
\usepackage{graphicx}
\usepackage{float}
\usepackage{placeins} 
\usepackage{booktabs}
\usepackage{iopams} 
\usepackage{hyperref}
\usepackage{fancyhdr}
\hypersetup{
    colorlinks=true,
    linkcolor=black, 
    citecolor=black,
    filecolor=black,
    urlcolor=black
}

\usepackage{acronym}
\acrodef{EMG}{electromyography}
\acrodef{HD-sEMG}{high-density surface electromyography}
\acrodef{MUAP}{motor unit action potential}
\acrodef{SCD}{Swarm-Contrastive Decomposition}
\acrodef{cBSS}{convolutive Blind Source Separation}
\acrodef{BSS}{blind source separation}
\acrodef{ICA}{Independent Component Analysis}
\acrodef{pps}{pulses per second}
\acrodef{MVC}{Maximum Voluntary Contraction}
\acrodef{RoA}{Rate of Agreement}
\acrodef{SNR}{signal-to-noise ratio}
\acrodef{IQR}{interquartile}
\acrodef{TA}{Tibialis Anterior}

\usepackage[style=apa, backend=biber]{biblatex}

\begin{document}

\fancyhf{}
\fancyhead[OC]{Unlocking High-Density Surface EMG}
\fancyhead[R]{\thepage}

\renewcommand{\headrulewidth}{0pt} 
\renewcommand{\footrulewidth}{0pt} 

\title[]{Unlocking the Full Potential of High-Density Surface EMG: Novel Non-Invasive High-Yield Motor Unit Decomposition}

\author{Agnese Grison$^{1}$, Irene Mendez Guerra$^{1}$, Alexander Kenneth Clarke$^{1}$, Silvia Muceli$^{1,2}$, Jaime Ib\'{a}\~{n}ez Pereda$^{1,3, 4}$ and Dario Farina$^1$}

\address{$^1$ Department of Bioengineering,
Imperial College London, London, UK}
\address{$^2$ Department of Electrical Engineering,
Chalmers University of Technology, Gothenburg, Sweden}
\address{$^3$ BSICoS Group in Aragón Institute of Engineering Research (I3A) and Instituto de Investigación Sanitaria de Aragón (IIS Aragón), Zaragoza, Spain }
\address{$^4$ Centro de Investigación Biomédica en Red en Bioingeniería, Biomateriales y Nanomedicina (CIBER-BBN), Zaragoza, Spain }

\ead{d.farina@imperial.ac.uk}
\vspace{10pt}
\begin{indented}
\item[]October 2024
\end{indented}

\begin{abstract}
The decomposition of high-density surface electromyography (HD-sEMG) signals into motor unit discharge patterns has become a powerful tool for investigating the neural control of movement, providing insights into motor neuron recruitment and discharge behavior. However, current algorithms, while very effective under certain conditions, face significant challenges in complex scenarios, as their accuracy and motor unit yield are highly dependent on anatomical differences among individuals. This can limit the number of decomposed motor units, particularly in challenging conditions. To address this issue, we recently introduced Swarm-Contrastive Decomposition (SCD), which dynamically adjusts the separation function based on the distribution of the data and prevents convergence to the same source. Initially applied to intramuscular EMG signals, SCD is here adapted for HD-sEMG signals. We demonstrated its ability to address key challenges faced by existing methods, particularly in identifying low-amplitude motor unit action potentials and effectively handling complex decomposition scenarios, like high-interference signals. We extensively validated SCD using simulated and experimental HD-sEMG recordings and compared it with current state-of-the-art decomposition methods under varying conditions, including different excitation levels, noise intensities, force profiles, sexes, and muscle groups. The proposed method consistently outperformed existing techniques in both the quantity of decoded motor units and the precision of their firing time identification. For instance, under certain experimental conditions, SCD detected more than three times as many motor units compared to previous methods, while also significantly improving accuracy. These advancements represent a major step forward in non-invasive EMG technology for studying motor unit activity in complex scenarios.
\end{abstract}

\section{Introduction}
The nervous system regulates muscle force by transmitting signals from alpha motor neurons in the spinal cord to muscle fibers. The collective activity of these motor neurons, known as the neural drive to the muscle, determines the overall muscle activation and force production \parencite{de1994common, farina2015common}. Each motor neuron's discharge triggers a \ac{MUAP} in the muscle fibers it innervates \parencite{kandel2000principles}. The summation of \ac{MUAP}s from all active motor units generates the \ac{EMG} signal, which represents the electrical activity of the muscle during a contraction \parencite{de1979physiology}. Therefore, the \ac{EMG} signal comprises both a neural component — the motor neuron discharge times — and a peripheral/muscle component — the \ac{MUAP} waveforms \parencite{stashuk2001emg}. \ac{EMG} decomposition aims to separate these two components, enabling the precise isolation and extraction of motor neuron activity. 

For nearly a century, the primary method for investigating motor units has been through invasive \ac{EMG} techniques, involving the decomposition of recordings from needle or wire electrodes  \parencite{adrian1929discharge, farina2024neural}. These recordings have a high degree of selectivity, typically allowing the decoding of only a few motor units at a time, which limits their scalability and generalizability. As a result, much of our current understanding of human motor neuron function is derived from small, controlled datasets obtained from a narrow set of motor units per subject. While these studies have been critical in establishing foundational principles, they do not translate well to larger populations or more varied conditions. Surface \ac{EMG} (sEMG) is less selective compared to invasive \ac{EMG} and therefore traditionally lacked the precision required to isolate individual motor units. 

The challenge of decomposing poorly selective s\ac{EMG} recordings has been partially addressed by increasing the number of recording sites (electrodes) and applying \ac{BSS} techniques. Increasing electrode counts led to high-density surface EMG (\ac{HD-sEMG}) \parencite{merletti2019tutorial} and provided multiple observation points for the motor unit activities. Most \ac{BSS} methods use the principles of \ac{ICA} to separate sources, i.e., the motor units, from the observed mixtures. This is achieved through the use of a contrast function, a nonlinear measure of non-Gaussianity of the signals, that is applied to the estimated spike trains in an iterative fashion \parencite{hyvarinen2000independent}. 

Because the \ac{HD-sEMG} signals are convolutive mixtures of the motor neuron spike trains, decomposition of these signals require \ac{cBSS}. In \ac{cBSS}, spike trains are extracted by optimizing a contrast function that maximizes an approximation of the skewness or kurtosis of the sources, thereby enhancing their statistical independence.

While \ac{BSS} of \ac{HD-sEMG} has deeply improved the study of motor units using non-invasive recordings, several challenges remain in its practical application. One major issue is that the number of successfully decomposed motor units can vary greatly across conditions, muscles, and individuals \parencite{del2020tutorial}. This variability stems from the differences across conditions in the distribution of \ac{MUAP}s at the skin surface, which is an inherent limitation of surface recordings. Indeed, in some scenarios, \ac{HD-sEMG} decomposition may not be feasible at all due to these challenges. As a result, non-invasive motor unit investigations often focus on a narrow set of muscles, specific experimental subjects, and controlled recording conditions \parencite{de2022neural}.

The underlying reason for poor \ac{HD-sEMG} decomposition is that, in certain cases, the \ac{MUAP}s from different motor units exhibit highly similar waveforms in both space and time, making it difficult for conventional contrast functions used in \ac{BSS} to isolate them. This similarity limits the effectiveness of traditional decomposition algorithms. A promising direction for overcoming this limitation is the development of methods that enhance the differences between \ac{MUAP}s by implementing novel strategies for blindly determining the optimal contrast function during decomposition. These approaches could improve the differentiation of similar \ac{MUAP}s, leading to an increased number of decomposed motor units, improved classification accuracy, and broader applicability of s\ac{EMG} decomposition in more challenging scenarios.

In a recent study \parencite{grison2024particle}, we introduced \ac{SCD} for decomposing multi-channel intramuscular \ac{EMG} recordings. The method optimizes the contrast function dynamically for each source, increasing the separation between sources. Here, we use the concepts of \ac{SCD} and we adapt it to the decoding of surface recordings. 
By dynamically adapting the contrast function, we ensure that it is specifically tailored to the unique characteristics of the source being decomposed, significantly enhancing source separation performance. The adaptability of the contrast function distinguishes \ac{SCD} from other state-of-the-art methods, which typically fix the contrast function to polynomial approximations of the kurtosis or the skewness \parencite{holobar2003surface, negro2016multi}. Moreover, \ac{SCD} implements a peel-off strategy for sequential source removal, which allows the algorithm to detect smaller or more subtle sources that would have otherwise been overshadowed by the larger, more dominant ones.

The proposed method was extensively validated with both simulated signals and experimental recordings across multiple conditions, including varying excitation levels, noise intensities, \ac{MUAP} overlap, muscle groups, and sexes. In all analyses, \ac{SCD} was compared against current \ac{cBSS} approaches, and specifically the method proposed by Negro et al. \parencite*{negro2016multi}, which is a current representative state-of-the-art decomposition approach. Henceforth, in the following we will refer to \ac{cBSS} as to this particular method, representing one of the several solutions proposed to address the general convolutive problem. The results demonstrated a substantial increase in performance of \ac{SCD} in all conditions tested, proving the effectiveness of the approach in increasing both the number and accuracy of the decomposed motor units. These findings indicated that, while the new method does not fully eliminate all challenges, \ac{SCD} marks a significant advancement in \ac{HD-sEMG} decomposition. By increasing the yield and accuracy of decomposition, it broadens the recording conditions and experimental scenarios where reliable motor unit activity can be extracted, ultimately increasing the practical utility and robustness of surface \ac{EMG} for a wide variety of applications.

\section{Methods}
\subsection{EMG generation model}

The \ac{EMG} signal can be modeled using a linear time-invariant multiple-input–multiple-output system, which can be compactly expressed in matrix form as:

\begin{equation}
    {\bf{x}}(t) = \sum_{l=0}^{L-1}{\bf{H}}(l){\bf{{s}}}(t-l) + {\boldsymbol{\xi}}(t)
\end{equation}

where ${\bf{x}}(t)=\left[x_{1}(t), x_{2}(t), \ldots, x_{M}(t)\right]^{T}$ is the vector of $M$ \ac{EMG} channels, ${\bf{{s}}}(t)=\left[s_{1}(t), s_{2}(t), \ldots, s_{N}(t)\right]^{T}$ represent the $N$ motor unit spike trains and $\boldsymbol{\xi}(t)$ accounts for the additive noise. The matrix ${\bf{H}}(l)$ has size $M \times N$ and contains the $l^{th}$ sample of the \(L\)-sample long \ac{MUAP} waveforms that appear for each of the \(N\) motor units across \(M\) channels, assuming constant shape under stationary conditions.

The convolutional model can be transformed into an instantaneous mixture by augmenting the vector of sources to include the \(N\) original sources and their respective delayed versions. \(L\) represents the duration of the impulse response of the filter, which models the volume conductor. To ensure a favorable ratio between the number of observations, i.e. the signals recorded at each electrode location, and sources, the observations are also extended by 
\(R\) delayed versions. This reformulated instantaneous model is expressed as:

\begin{center}

\begin{equation} \label{eqn2}
\tilde{\mathbf{x}}(t) = \tilde{\mathbf{H}} \tilde{\mathbf{s}}(t) + \tilde{\boldsymbol{\xi}}(t)
\end{equation}
\end{center}

where  
\begin{center}
\begin{eqnarray}
\tilde{\mathbf{s}}(t) & = & \left[\tilde{\boldsymbol{s}}_1(t), \tilde{\boldsymbol{s}}_2(t), \ldots, \tilde{\boldsymbol{s}}_j(t), \ldots, \tilde{\boldsymbol{s}}_N(t)\right]^T \nonumber \\
\tilde{\boldsymbol{s}}_j(t) & = & \left[s_j(t), s_j(t-1), \ldots, s_j(t-(L+R+1))\right]
\end{eqnarray}
\end{center}
and $\tilde{\boldsymbol{H}}$ is constructed from the extended convolution kernels $\boldsymbol{\tilde{h}}$. The observed signal $\tilde{\boldsymbol{x}}(t)$ is:

\begin{center}
    
\begin{eqnarray}
\tilde{\mathbf{x}}(t) & = & \left[\tilde{\boldsymbol{x}}_1(t), \tilde{\boldsymbol{x}}_2(t), \ldots, \tilde{\boldsymbol{x}}_i(t), \ldots, \tilde{\boldsymbol{x}}_M(t)\right]^T \nonumber \\
\tilde{\boldsymbol{x}}_i(t) & = & \left[x_i(t), x_i(t-1), \ldots, x_i(t-R)\right]
\end{eqnarray}
\end{center}

In the absence of noise, the retrieval of discharge timings \(\mathbf{\tilde{s}}(t)\) can be framed as the following inverse problem:

\begin{eqnarray}
    \tilde{\mathbf{s}}(t) = \mathbf{W}\mathbf{\tilde{x}}(t)
\end{eqnarray}
where \(\mathbf{W}\) represents the approximate pseudo-inverse of \(\mathbf{\tilde{H}}\). The goal of the decomposition process is to determine the \(N\) separation vectors forming the matrix \(\mathbf{W}\). 

\subsection{Decomposition}
We propose the use of the \ac{SCD} decomposition algorithm \parencite{grison2024particle} and compare it with a representative state of the art approach for \ac{cBSS} decomposition \parencite{negro2016multi}. 

Both algorithms rely on \ac{ICA} to separate sources by maximizing the sparseness of the estimated sources \parencite{hyvarinen2000independent}. The non-linear function used to assess sparseness is called the contrast function, \(G\). Similar to the Gaussianity property, mixing sources results in a signal less sparse than the individual sources. Thus, sparse sources can be identified by finding projections that maximise sparseness, making the choice of contrast function critical for the stability of the numerical optimization process. Cumulants are a widely used class of non-linearities for this purpose. By optimizing the separation vector to maximize higher-order cumulants, such as skewness or kurtosis, non-Gaussianity as well as sparseness are amplified, increasing the likelihood of isolating sparse sources.

In the \ac{cBSS} methods used for comparison \parencite{negro2016multi}, the contrast function is defined to maximize an approximation of the third cumulant, i.e., the skeweness, of the sources, utilizing \(G(s) = \frac{1}{6}\mathrm{y^3}\). 
In contrast, our approach proposes to maximize an adaptively tuned higher-order cumulant for each source. Specifically, \(G(s) = E \{sign(s)|s|^e \}\), where \(e\) represents the exponent of the polynomial function, and \(E\) is the expectation operator. This is the main difference between \ac{cBSS} and the algorithm \ac{SCD}. 

In \ac{SCD}, candidate separation vectors \(\mathbf{w}\) were randomly initialized from a zero-mean normal distribution. \ac{ICA} was run for a maximum of 1000 iterations, with an early termination criterion applied if, after 20 iterations, the loss \(G(s)\) failed to improve. After each \ac{ICA} step, a peak-finding algorithm detected the source samples, followed by a two-class \(k\)-medoid clustering to identify potential motor unit spikes. Source quality was evaluated using a fitness function based on the coefficient of variation of the interspike intervals, where the candidate with the lowest coefficient of variation was selected as the best source. The coefficient of variation was chosen as an appropriate metric due to the regularity of motor unit firing in isometric contractions.

After each \ac{ICA} update, the optimal exponent for the contrast function was chosen from the current pool of candidates and updated by moving toward the exponent that yielded the highest-quality source. The separation vectors were then reinitialised with the spike-triggered-average of the highest-quality source. Optimization was halted after 10 updates of the optimal exponent coefficient. To assess the final source quality, the silhouette measure \parencite{negro2016multi} was used to evaluate the clustering separation between the decomposed source and the background noise. If the source passed this evaluation (silhouette \textgreater0.85) and had not been identified in previous iterations, it was peeled off from the signal to prevent further convergence in subsequent decomposition iterations. However, from the estimated discharge times, motor units exhibiting discharge properties outside the predefined thresholds—specifically, a coefficient of variation above 35\% or a firing rate outside the 2-35 Hz range \parencite{martinez2017tracking}—were excluded from further analysis. 

To assess the importance of the two core features of \ac{SCD} — the adaptive contrast function and the peel-off procedure — two targeted ablations were conducted. The first ablation focused on addressing repeated convergence to the same source. Three strategies were tested: (1) initializing the separation vectors \(\mathbf{w}\) with the activity index, a proxy for global pulse train activity \parencite{holobar2007gradient}, and eliminating the peel-off process, while preventing previously accepted motor unit firing timings from being reinitialized by excluding them from the activity index; (2) combining the activity index with a source deflation method to enforce orthogonality between separation vectors; and (3) comparing with the proposed \ac{SCD}.

The second ablation maintained the peel-off method across all configurations and examined the effect of different exponents in the contrast function. These were compared against the performance of \ac{SCD} with the dynamically adapted exponents.

\subsection{Simulations}
\ac{HD-sEMG} data was simulated using NeuroMotion \parencite{ma2024neuromotion}, an advanced \ac{EMG} simulator designed to produce physiological electric potentials during voluntary forearm movements. NeuroMotion operates through three modules. The first, an upper-limb musculoskeletal model developed using the OpenSim API \parencite{delp2007opensim}, defines and visualizes movements while estimating muscle fibre lengths and muscle activation levels. These estimates are then input into BioMime \parencite{ma2024conditional}, an Artificial Intelligence-based volume conductor model that generates \ac{MUAP}s based on parameters like fibre number, depth, angular position, innervation zone, and conduction velocity, derived from a myoelectric digital-twin model \parencite{maksymenko2023myoelectric}. The final module, a motor unit pool model, converts neural inputs into spike train simulations, completing the muscle activation simulation \parencite{fuglevand1993models}. 

An isometric index finger contraction from the flexor digitorum superficialis was simulated with NeuroMotion. The simulated signals were generated for an electrode band comprising a 10 × 32 electrode grid, positioned around the proximal third of the forearm.
Figure \ref{fig:setup} \textbf{a} displays a schematic of the simulated setup. A pool of 100 motor neurons  was simulated. The innervated muscle fibers were randomly and uniformly distributed throughout the muscle volume, with innervation zones normally distributed around 50\%±10\% of the total fiber length. An exponential function was employed to model the recruitment thresholds and number of innervated fibers for each motor neuron, resulting in a higher proportion of small, low-threshold motor neurons compared to larger, high-threshold ones.
Conduction velocities were sampled from a normal distribution (mean 4.0±0.5 ms$^{-1}$) and sorted based on the number of innervated fibers. Motor neurons  began firing at a baseline rate of 8 \ac{pps} once the excitation level surpassed their recruitment threshold  \parencite{fuglevand1993models}. The discharge rate increased linearly by 3 \ac{pps} for every 10\% increase in excitation \parencite{fuglevand1993models, keenan2006amplitude}. Maximum discharge rates varied from 35 \ac{pps} for the first recruited motor neuron to 25 \ac{pps} for higher-threshold motor neurons  \parencite{fuglevand1993models}. The final motor neuron was recruited at 50\% of the maximum excitation level \parencite{fuglevand1993models}. The variability in discharge times followed a Gaussian random process with a coefficient of variation of 0.2 \parencite{fuglevand1993models, keenan2006amplitude}.
While the myoelectric digital-twin model \parencite{maksymenko2023myoelectric} provided the muscle geometry based on magnetic resonance imaging, NeuroMotion estimated other motor unit parameters (position, length, depth, angle, innervation zone, and conduction velocity) based on physiological ranges. These distributions were randomized over 10 bootstrapping iterations to capture the full variability observed in real data.
Three distinct analyses were performed on different types of simulated data to investigate the characteristics of the proposed algorithm. The first analysis examined the effect of excitation level on the decomposition process by gradually increasing the excitation from 10\% to 100\% of \ac{MVC} in 10\% increments, forming a dataset of 100 recordings, given by 10 bootstrapping iterations for each of the 10 force levels. Each contraction lasted 30 s. Throughout this analysis, the \ac{SNR} was maintained at 25 dB (Gaussian additive noise model) to isolate the effect of the excitation level. 

The second analysis investigated the impact of the noise level on motor unit decomposition. In these simulations, the excitation level was fixed at 30 \%\ac{MVC} during 30-second contractions, while the noise level was varied from 10 to 30 dB, with 5-dB increments. This approach allowed for the examination of the algorithm performance under different levels of signal contamination, simulating more challenging real-world conditions.
The last analysis focused on ballistic contractions (high synchronization level), where the force increased rapidly from 0 to 40 \%\ac{MVC}. Thirty bursts of muscle activity were simulated, each lasting 1 second and separated by 3-second intervals. These conditions mimicked sudden, explosive muscle contractions, enabling the evaluation of the algorithm performance under high levels of synchronised motor unit activity. The noise level in this experiment was maintained at 25 dB. It is noteworthy that while the contraction was dynamic in terms of force, the posture and  muscle geometry remained constant throughout the simulation.

\subsection{Experimental Data}
\begin{figure}[!ht]
    \centering
    \includegraphics[ width=\columnwidth]{ 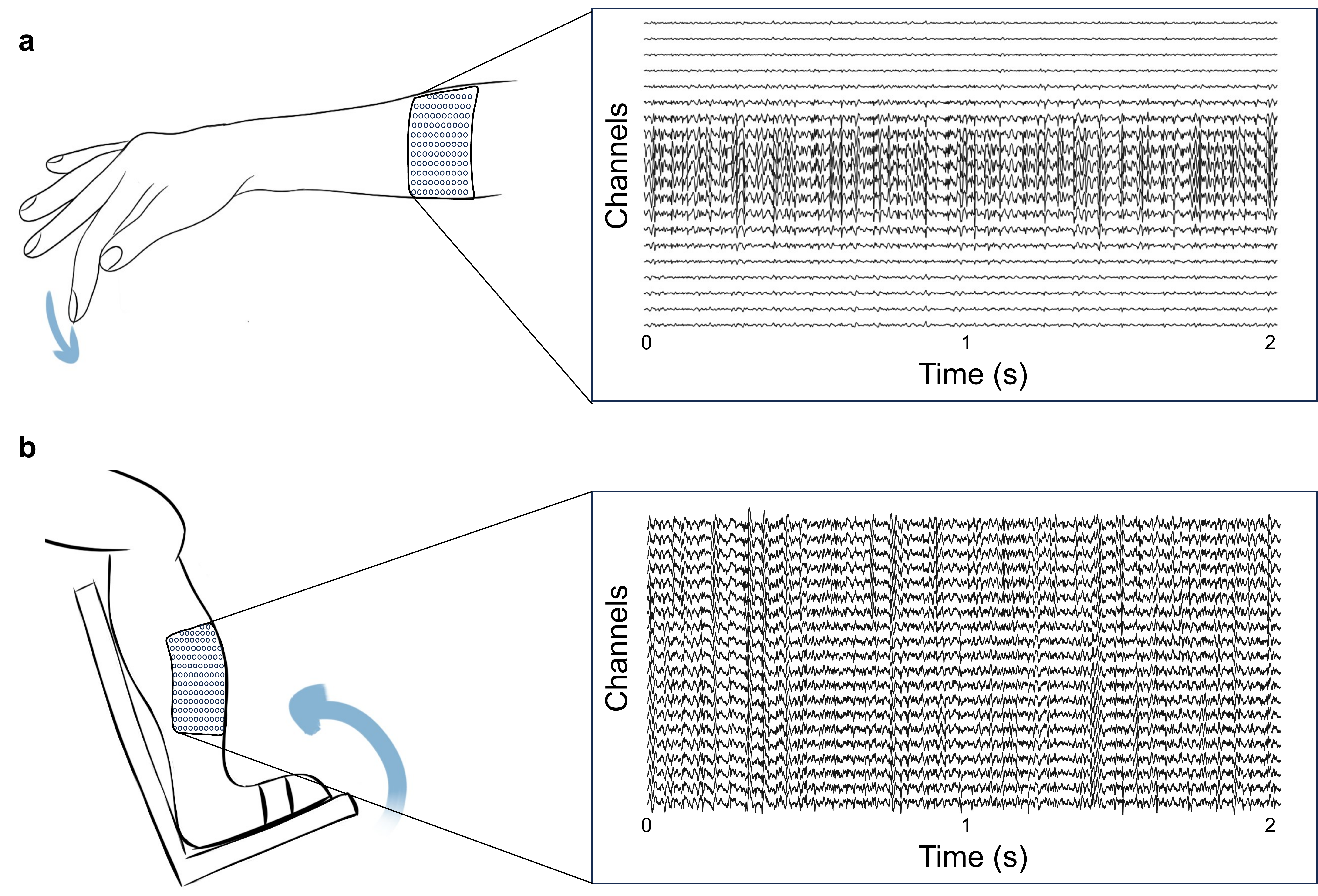}
    \caption{Schematics of the \ac{HD-sEMG} data used in the analysis. \textbf{a} Simulated \ac{HD-sEMG} data from forearm muscles during index finger flexion, with the electrode grid positioned over the proximal third of the forearm. The same setup was applied for the experimental recordings from the forearm. Representative data (simulations) is shown for a 30 \%MVC contraction. \textbf{b} Experimental \ac{HD-sEMG} data were also recorded from the \ac{TA} muscle during ankle dorsiflexion. Representative data for a 20 \%MVC contraction is displayed.}
    \label{fig:setup}
\end{figure}

The experiments adhered to the ethical guidelines set by Imperial College London (ICREC Project ID 19IC5640). All procedures were conducted in accordance with the Declaration of Helsinki, with informed consent obtained from the participants prior to each experiment.
Three experiments were conducted, with the first two focusing on the \ac{TA} muscle of healthy male and female participants respectively, and the last one focusing on the forearm of healthy female participants. In the \ac{TA} experiments, participants were seated with their right leg and foot secured to a dynamometer and they were instructed to sustain an isometric ankle dorsiflexion. Figure \ref{fig:setup} \textbf{b} displays a schematic of the experimental setup. Similar to the simulated conditions, subjects in the forearm experiments were instructed to perform isometric index finger flexion (Figure \ref{fig:setup}, \textbf{a}). The force levels were set as percentages of the participants' maximal voluntary contraction (MVC), with visual feedback provided for both exerted force and target. \ac{EMG} signals were recorded in monopolar derivation, with a reference electrode placed on the ankle (\ac{TA}), and wrist (forearm), and bandpass-filtered between 20 and 500 Hz. All signals and force data were recorded concurrently using a multi-channel amplifier (OT-Bioelettronica, Torino, Italy), high-pass filtered at 10 Hz, and digitized at 16-bit resolution.

\subsubsection{Two-source validation (TA)}
Two healthy men, aged 39 and 30, were recruited. The protocol included 20-second isometric contractions at 10, 20, and 30 \%\ac{MVC}, a 15-second contraction at 40 \%\ac{MVC}, and 10-second contractions at 50, 60, and 70 \%\ac{MVC}. Three 40-channel HD-i\ac{EMG} micro-electrode arrays \parencite{muceli2022blind, muceli2015} were implanted in the \ac{TA}, oriented longitudinally and spaced approximately 3-cm apart. Two 64-channel \ac{HD-sEMG} grids (4 mm inter-electrode distance, 13x5 electrode configuration) were placed on the skin surface above the intramuscular detection sites. Intramuscular and surface \ac{EMG} signals were concurrently sampled at 10,240 Hz. The concurrent recording of intramuscular and surface EMG signals allowed for the use of the two-source validation methods for an objective and rigorous assessment of decomposition accuracy \parencite{farina2014extraction, mambrito1984technique}.

\subsubsection{Sex differences in motor unit yield (TA)}
Two healthy women, aged 23 and 24, were recruited. The protocol included 20-second isometric contractions at 5 and 10 \%\ac{MVC}. One 256-channel \ac{HD-sEMG} grid (4 mm inter-electrode distance, 32x8 electrode configuration) was placed on the \ac{TA}. \ac{HD-sEMG} signals were sampled at 2,048 Hz. The measurements on female individuals allowed to compare the algorithms in conditions that are challenging for surface EMG decomposition since it has been reported that decomposition yield and accuracy decreases in female individuals \parencite{taylor2022sex, lulic2022sex, del2020tutorial}.

\subsubsection{Motor unit yield in complex muscle groups (forearm)}
Two healthy women, aged 26 and 30, were recruited for this study. The protocol included 20-second isometric index finger flexions at 15 \%\ac{MVC}. Three 64-channel \ac{HD-sEMG} grids (8 mm inter-electrode distance, 13x5 electrode configuration) were placed around the proximal third of the forearm of each participant. \ac{HD-sEMG} signals were sampled at 2,048 Hz. The measurements on the forearm allowed to analyze the performance of the algorithms in more complex muscle groups, where the number of motor units decomposed is usually limited to less than ten \parencite{del2020tutorial}.

\subsection{Metrics for accuracy}
The level of agreement between the discharge times of motor units as decoded from the decomposition of the simulated \ac{HD-sEMG} and the ground truth was assessed using the \ac{RoA}. The \ac{RoA} measures the fraction of commonly identified discharges relative to the total number of discharges, considering both common and not-common discharge times. The \ac{RoA} was therefore calculated as follows:

\[RoA = \frac{TP}{TP + FP_1 + FP_2}\]

where \(TP\) refers to the number of matched predicted activations within a deviation margin of ±0.5 ms \parencite{farina2001evaluation}. \(FP_1\) and \(FP_2\) represent the counts of unmatched predicted activations, corresponding to discharge times present in only one of the two sets. 

Several methods have been proposed to assess the accuracy of decomposition of experimental \ac{HD-sEMG} data \parencite{holobar2010experimental, holobar2014accurate, hu2013assessment}, with the two-source validation being the most reliable \parencite{mambrito1984technique, farina2014extraction}. This approach involves recording both \ac{HD-sEMG} and HD-i\ac{EMG} signals concurrently, decomposing them independently, and comparing the discharge times from the two decompositions to determine the \ac{RoA}, as defined above. This approach operates on the principle that similar results from two independent algorithms applied to different signals are likely correct, as the probability of identical errors is low. The \ac{RoA} thus reflects the relative performance of the algorithms without bias toward either method \parencite{farina2014extraction}. Here, when available, we utilized the concurrent recording of the HD-i\ac{EMG} micro-electrode arrays to validate the surface decomposition. The HD-i\ac{EMG} data was decomposed using \ac{SCD}, as described in  \parencite{grison2024particle}. When the concurrent recording of the HD-i\ac{EMG} signals was not available, the number of motor units decoded by \ac{SCD} and \ac{cBSS} was compared, and the \ac{RoA} between the commonly identified motor units (by \ac{SCD} and \ac{cBSS}) was also analyzed.

For all analyses, the \ac{RoA} is reported in percentages as the median with the \ac{IQR} ranges in brackets.

\subsection{Statistical Analysis}
All statistical analyses were conducted in Python 3.10 using the SciPy library \parencite{2020SciPy}, with a significance threshold set at \(p\)\textless0.05. Non-parametric tests were applied when normality could not be verified, as determined by the Shapiro–Wilk test (\(p\)\textless0.05). Significance levels are denoted as \emph{ns} non-significant, *** \(p\)\textless0.001, ** \(p\)\textless0.01, and *\(p\)\textless0.05.

To compare the number of decomposed motor units, the \ac{RoA} between decomposed motor units and the simulated ground truth, and various motor unit characteristics, Mann–Whitney U tests were employed. The paired Wilcoxon signed-rank test was used to evaluate the \ac{RoA} of motor units that were matched with the HD-i\ac{EMG} recordings and commonly identified by both \ac{cBSS} and \ac{SCD}.

\section{Results}
\subsection{Simulations: Excitation level} \label{exp_25db}

\begin{figure}[!htbp]
    \centering
    \includegraphics[ width=\columnwidth]{ 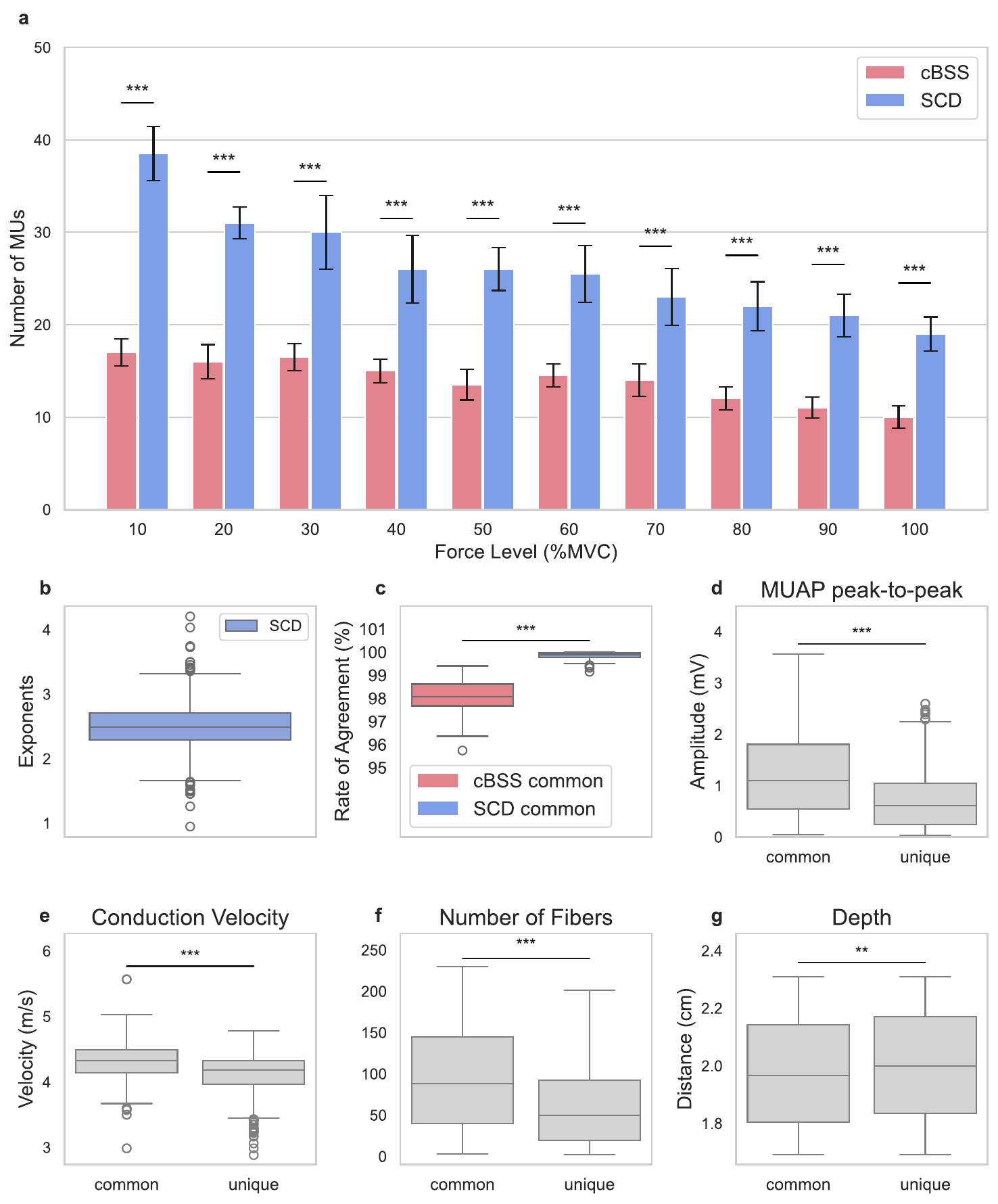}
    \caption{\textbf{a} Median number of motor units per bootstrap iteration for \ac{SCD} (blue) and \ac{cBSS} (pink). \textbf{b} Distribution of exponents for \ac{SCD}. \textbf{c} Distribution of the \ac{RoA} between the automatic methods and the simulated ground truth for the motor units commonly identified by \ac{SCD} and \ac{cBSS}.  \textbf{d-g} Distributions of the peak-to-peak \ac{MUAP} amplitudes (\textbf{d}), conduction velocity (\textbf{e}), number of fibers innervated (\textbf{f}), and depth (measured with respect to the skin, higher is deeper) (\textbf{g}) for motor units common to both \ac{SCD} and \ac{cBSS}, and those uniquely identified by \ac{SCD}.}
    \label{fig:static}
\end{figure}

Figure \ref{fig:static} summarizes the results of motor unit decomposition using \ac{SCD} and \ac{cBSS} across varying excitation levels in the simulated data. Panel \textbf{a} illustrates the median number of decomposed motor units over 10 bootstrapping iterations. Across all excitation levels and bootstrapping iterations, on average, \ac{cBSS} identified 13.9±2.7 motor units per contraction, while \ac{SCD} identified on average nearly a double number of motor units (25.9±5.8 motor units), which was a statistically significant increase (\(p\)\textless0.001). Of the motor units identified by \ac{cBSS}, 98\% were also detected by \ac{SCD}, demonstrating that \ac{SCD} primarily expanded the total count of detected units, rather than identifying significantly different units from those already captured by \ac{cBSS}. Figure \ref{fig:static} \textbf{b} reports the final exponent distribution of the automatically selected contrast functions for all identified sources with \ac{SCD}. The median \ac{RoA} of all the motor units decomposed by \ac{SCD} was 100\% (\ac{IQR} 98.8\% - 100\%), compared to 99.4\% (\ac{IQR} 96.8\% - 100\%) for \ac{cBSS}. Figure \ref{fig:static} \textbf{c} shows the \ac{RoA} for the motor units identified by both methods. In these units, \ac{SCD} showed a significant increase in \ac{RoA} (\(p\)\textless0.001).

A comparison between the motor units which were detected by both \ac{SCD} and \ac{cBSS} (\textit{common}) and those identified only by \ac{SCD} (\textit{unique}) is presented across various motor unit properties. These properties include peak-to-peak amplitudes of the \ac{MUAP} in the channel with the highest amplitude (Figure \ref{fig:static}, \textbf{d}), motor unit conduction velocities (Figure \ref{fig:static}, \textbf{e}), the number of fibers innervated per motor unit (Figure \ref{fig:static}, \textbf{f}), and the depth of the motor units (Figure \ref{fig:static}, \textbf{g}). These results show that the motor units uniquely identified by \ac{SCD} tended to exhibit lower peak-to-peak amplitudes, suggesting that these motor units were either located farther from the electrodes, or deeper within the muscle. However, Figure \ref{fig:static} \textbf{g} shows that, while significant, the difference in distance from the electrodes between common and unique units was relatively small. Moreover, the \ac{SCD}-unique motor units typically had slower conduction velocities and a smaller number of innervated fibers, indicating they were smaller motor units. 

\subsubsection{Ablations}
\begin{figure}[H]
    \centering
        \includegraphics[ width=\columnwidth]{ 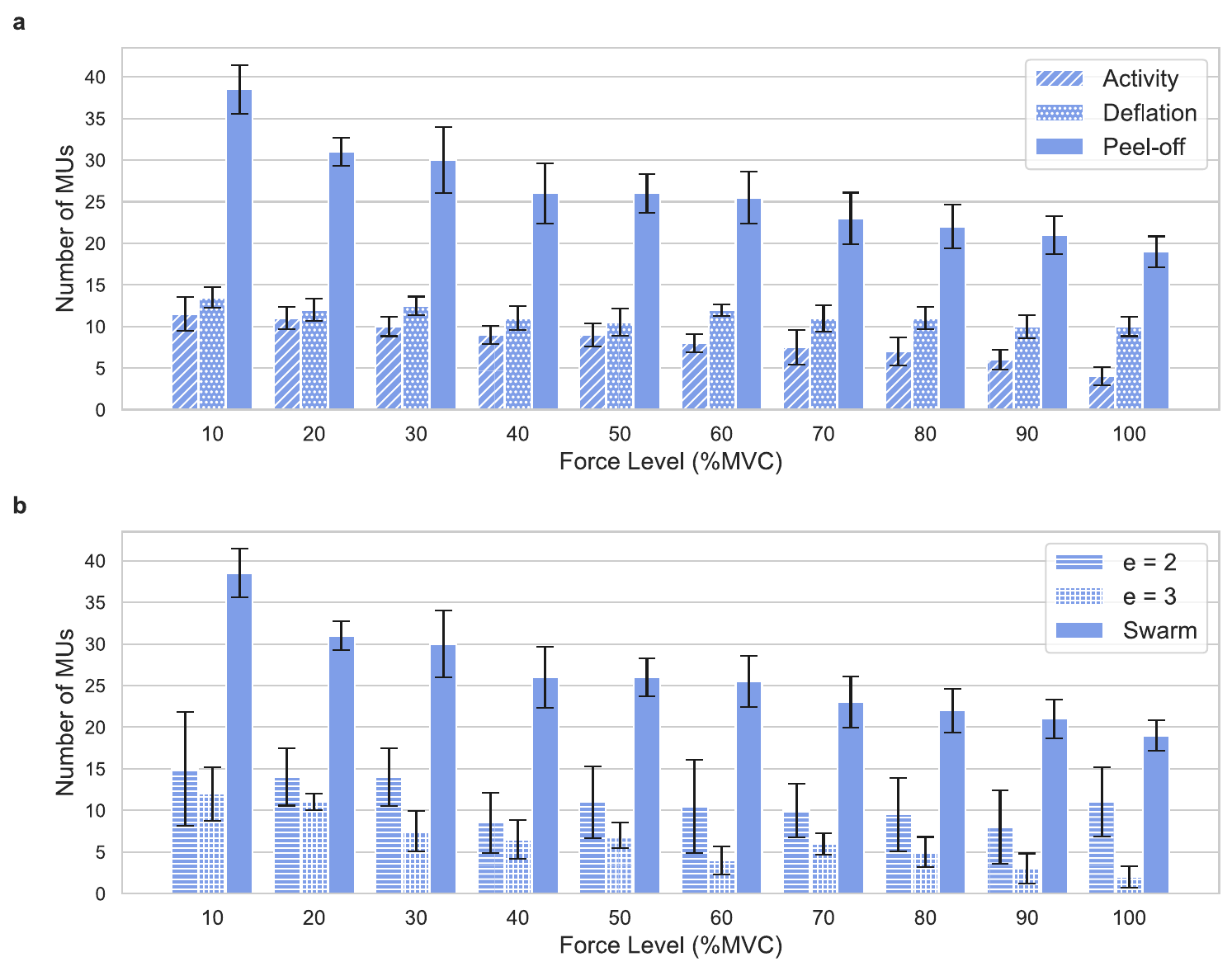}
    \caption{\textbf{a} Effect of the method used to prevent source convergence on the number of motor units found. Three methods are reported: 1) activity (use the activity index to initialize the separation vectors), 2) deflation (activity index to initialize the separation vectors and orthogonalize the separation vectors), 3) peel-off (remove found sources from the \ac{EMG}). \textbf{b} Effect of the exponent of the contrast function on the number of motor units found. Peel-off approach was used for all the three methods reported: 1) exponent fixed at 2, 2) exponent fixed at 3, 3) exponents starting at [2,3,4,5,6,7] and updated with particle swarm optimisation.}
    \label{fig:ablations}
\end{figure}

Figure \ref{fig:ablations} \textbf{a} displays the median number of decomposed motor units across bootstrapping iterations, stratified by force level, for the three methods employed to mitigate repeated convergence to the same source during the \ac{ICA} optimization. On average, across force levels and bootstrapping iterations, the activity index method identified 8.4±2.6 motor units, the deflation method 11.3±1.7 motor units, and the peel-off method 25.9±5.8 motor units. The results with the three methods were statistically different from each other (\(p\)\textless0.001).

To further understand the interaction between the ablation method and the way in which contrast functions are selected, we compared the number of units decomposed using the best ablation method in the previous test (peel-off) and three ways of defining the contrast functions for the separation: fixed contrast functions with exponents of orders 2 or 3, and using \ac{SCD} to optimize the selection of the contrast function used for each decomposed source (Figure \ref{fig:ablations}, \textbf{b}). On average, across all force levels and bootstrapping iterations, the method with $e=2$ identified 10.9±4.6 motor units, the method with $e=3$ identified 6.4±3.4 motor units, and the method with the adaptive exponent identified 25.9±5.8 motor units, which were all statistically different from each other (\(p\)\textless0.001). 

These results reflect an important improvement in the decomposition when an adaptive method to select contrast functions is combined with the peel-off method for iterative source removal. Using either of these two configurations alone with either fixed contrast functions or other possible source removal methods resulted in drastic reductions in the number of decomposed motor units.

\subsection{Simulations: Noise level}
\begin{figure}[H]
    \centering
    \includegraphics[ width=\columnwidth]{ 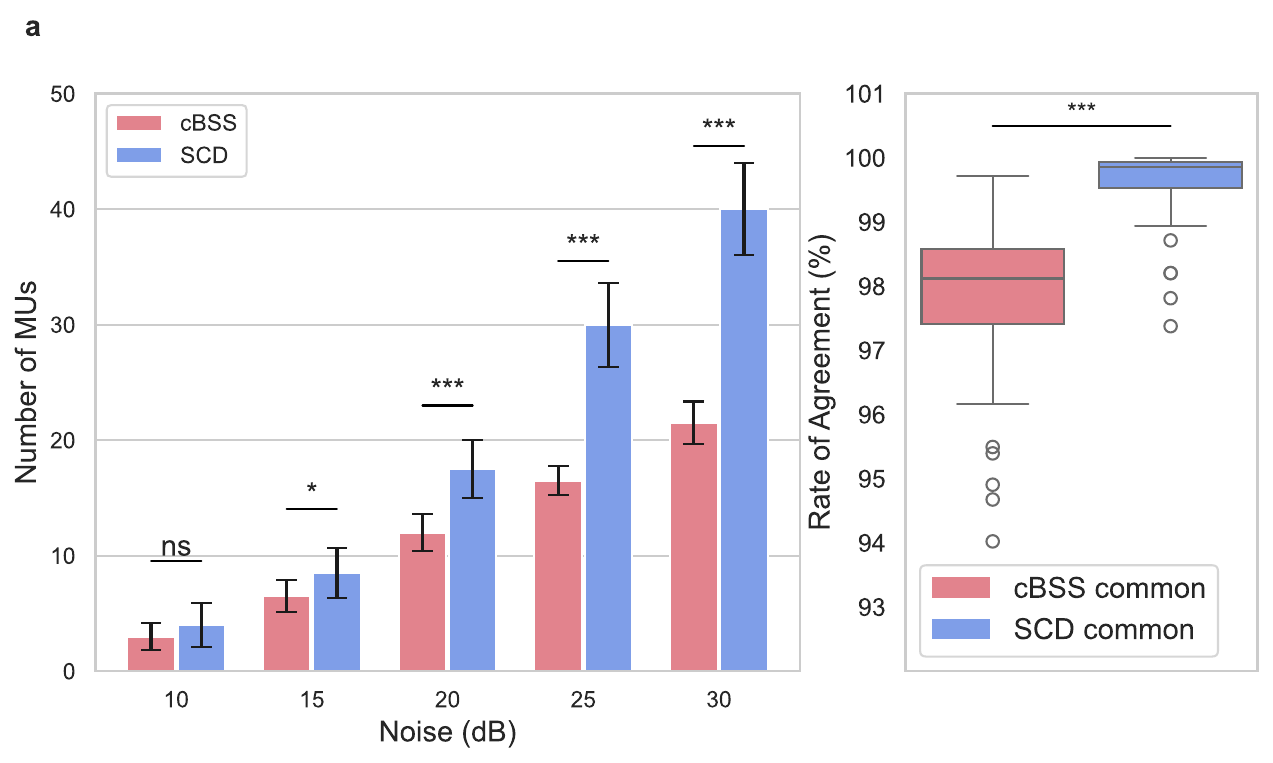}
    \caption{Effect of noise on the number of motor units found and the \ac{RoA} with the ground truth for 30 \%\ac{MVC} force level. \textbf{a} Number of motor units against noise level, for \ac{SCD} and \ac{cBSS}. \textbf{b} Distribution of the \ac{RoA} between decomposed motor units and their simulated ground truth for the motor units commonly identified by \ac{SCD} and \ac{cBSS}.}
    \label{fig:results_noise}
\end{figure}

Figure \ref{fig:results_noise} presents the median number of decomposed motor units across bootstrap iterations at a fixed excitation level of 30 \%\ac{MVC} (panel \textbf{a}) and the \ac{RoA} of the identified motor units compared to the simulated ground truth. \ac{SCD} consistently detected significantly more motor units across all noise levels, except at 10 dB, where the increase was not statistically significant (\(p\)=0.18). On average across all noise levels and bootstrapping iterations, \ac{cBSS} identified 11.9±6.9 motor units, while \ac{SCD} detected 19.8±13.5 motor units. Approximately 97\% (11.5±7.2) of the motor units detected by \ac{cBSS} were also detected by \ac{SCD}. The \ac{RoA} for the motor units identified by \ac{SCD} was 100\% (\ac{IQR} 99.2\% - 100\%), compared to 99.5\% (\ac{IQR} 97.0\% - 100\%) for \ac{cBSS}. The \ac{RoA} of the motor units commonly identified by both \ac{SCD} and \ac{cBSS} (Figure \ref{fig:results_noise}, \textbf{b}) was significantly higher for \ac{SCD} compared to \ac{cBSS} (\(p\)\textless0.001). 

\subsection{Simulations: Ballistic contractions} \label{exp_ballistic}
The decomposition performance was preserved in contractions simulated with ballistic force changes, where the overlap of the \ac{MUAP}s over time was greater due to the rapid recruitment and de-recruitment of the motor units. 

\begin{figure}[H]
    \centering
        \includegraphics[ width=\columnwidth]{ 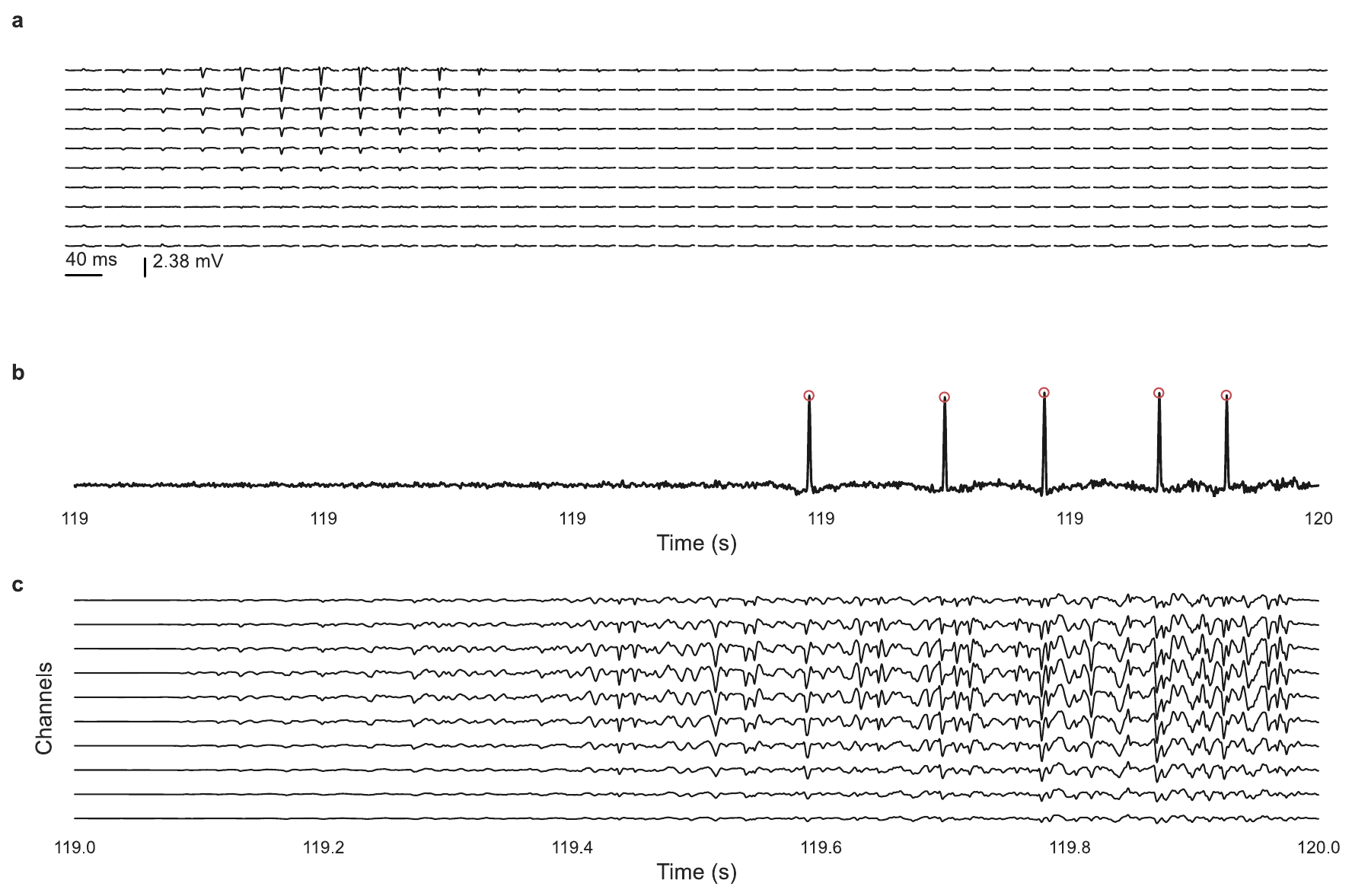}
    \caption{Representative example of a unit in a ballistic contraction decomposed with 100\% accuracy. \textbf{a} Spatial distribution of the \ac{MUAP} arranged in the 10x32 electrode configuration. \textbf{b} A one-second zoom-in of the innervation pulse train of the full source. The discharge times of the clustered source are shown in red. \textbf{c} A one-second zoom-in of the \ac{EMG}.}
    \label{fig:ballistic_source}
\end{figure}

Figure \ref{fig:ballistic_source} provides a representative example of the signal and the decomposed activity. Panel \textbf{a} presents the spike-triggered average of the MUAP of a decomposed unit. The estimated activity of the unit in one of the bursts of EMG is shown in panel \textbf{b}, which illustrates a clear separation between the source components (marked with red circles) and the background activity. A representative example of ten channels of the simulated \ac{EMG} for the same one-second interval is depicted in panel \textbf{c}, reported for the specific spatial grid configuration (10x32).

\begin{figure}[H]
    \centering
    \includegraphics[ width=\columnwidth]{ 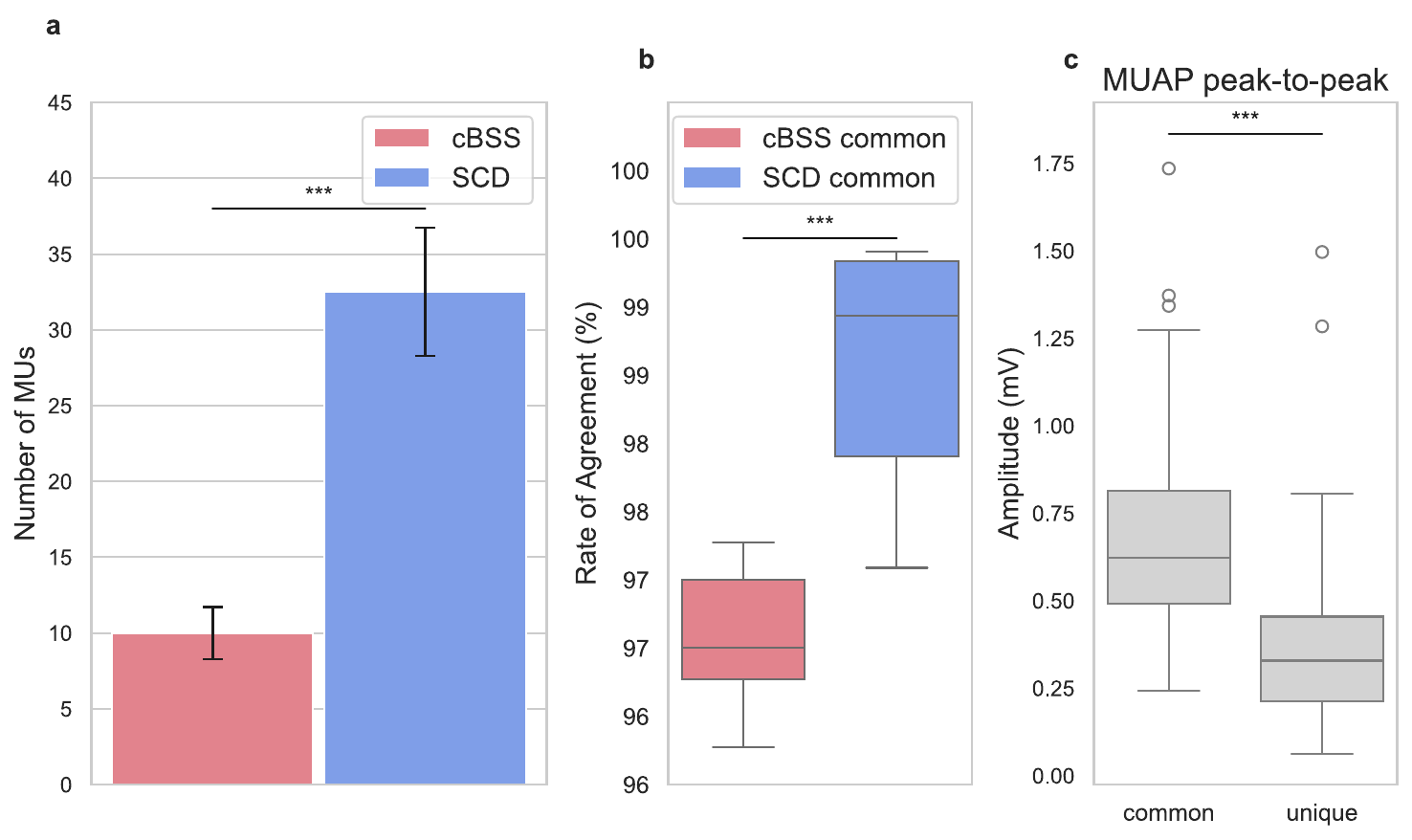}
    \caption{Effect of ballistic task on the number of motor units found and the \ac{RoA} with the ground truth. \textbf{a} Number of motor units found for \ac{SCD} and \ac{cBSS}. \textbf{b} Distribution of the \ac{RoA} between decomposed motor units and their simulated ground truth for the motor units commonly detected by \ac{SCD} and \ac{cBSS}. \textbf{c} Distributions of the peak-to-peak \ac{MUAP} amplitudes for motor units common to both \ac{SCD} and \ac{cBSS}, and those uniquely identified by \ac{SCD}.}
    \label{fig:results_ballistic}
\end{figure}

Figure \ref{fig:results_ballistic} presents the results of this analysis. Panel \textbf{a} shows the number of decomposed motor units for both \ac{cBSS} and \ac{SCD}, with \ac{SCD} identifying approximately three times the number of motor units of \ac{cBSS}. On average across the bootstrapping iterations, \ac{cBSS} detected 10.5±1.7 motor units, while \ac{SCD} decoded 31.2±4.3, which were statistically different (\(p\)\textless0.001). On average, 96\% of the motor units identified with \ac{cBSS} were also identified with \ac{SCD}. The \ac{RoA} of all the motor units decomposed by \ac{SCD} was 99.0\% (96.5\% - 100\%), while \ac{cBSS} achieved 97.5\% (96.2\% - 99.0\%). Out of the commonly identified motor units, \ac{SCD} achieved a significantly higher \ac{RoA} (\(p\)\textless0.001). The distribution of \ac{RoA} for the commonly detected motor units is illustrated in Figure \ref{fig:results_ballistic}, panel \textbf{b}. Additionally, the motor units uniquely identified by \ac{SCD} displayed significantly lower (\(p\)\textless0.001) peak-to-peak \ac{MUAP} amplitudes compared to those found by \ac{cBSS} (panel \textbf{c}). This further demonstrates \ac{SCD}'s capability to decompose lower-amplitude motor units that are missed by \ac{cBSS}.

\subsection{Experiments: Two-source validation}
This experiment aimed to compare the \ac{HD-sEMG} decomposition results between \ac{SCD} and \ac{cBSS} using experimental data. The outputs from both methods were validated by comparing them against the decomposition of concurrently recorded HD-i\ac{EMG} signals, providing a reliable benchmark for assessing the accuracy of the surface decompositions.

\begin{figure}[!ht]
    \centering
    \includegraphics[ width=\columnwidth]{ 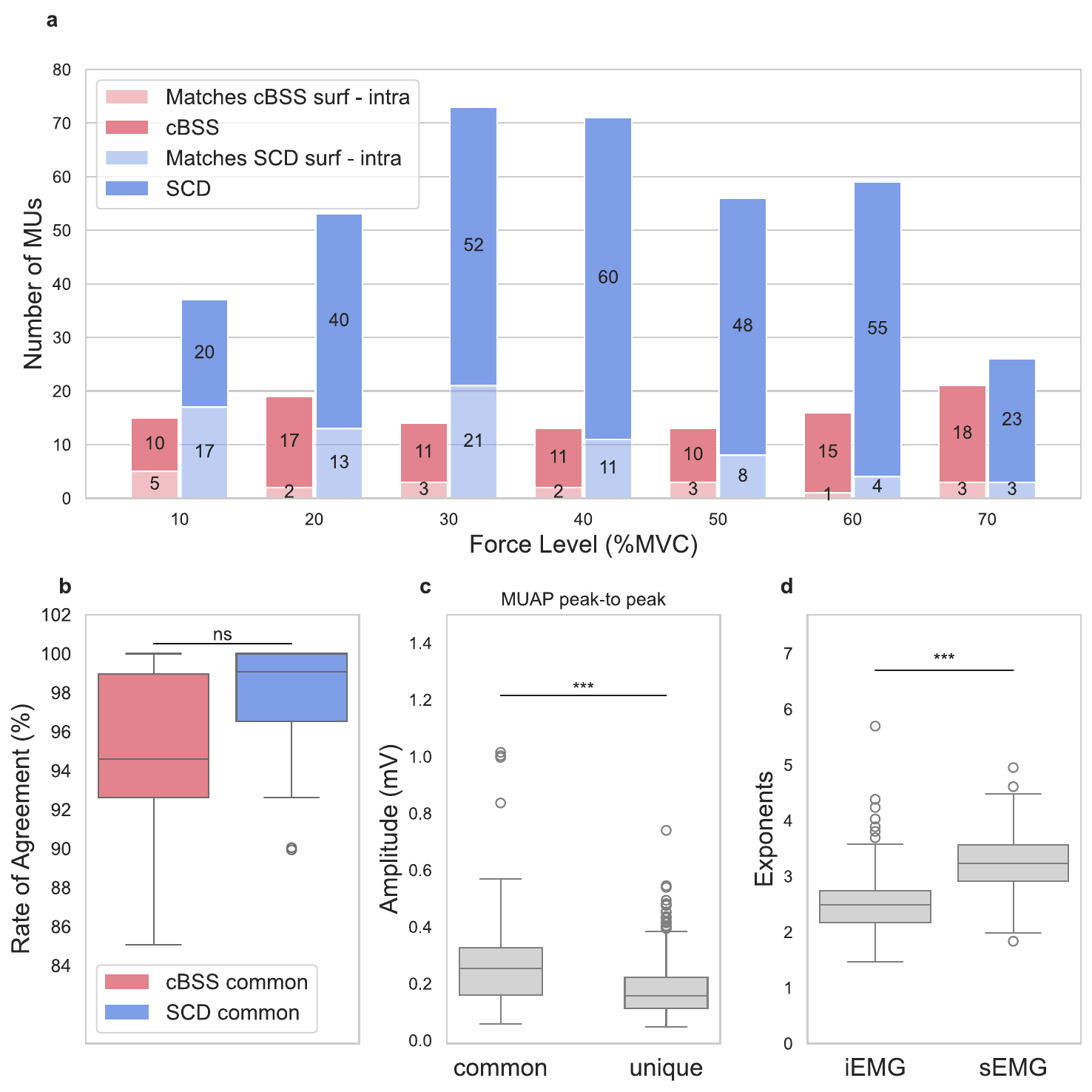}
    \caption{Effect of the decomposition algorithm used on experimental data. \textbf{a} Number of motor units found against the force level for \ac{SCD} and \ac{cBSS}. The number of matched motor units between the intramuscular and the surface recordings is reported in a lighter shade (light pink for \ac{cBSS} and light blue for \ac{SCD}). The results from the two subjects are pooled together. \textbf{b} Distribution of the \ac{RoA} for the motor units that matched with the intramuscular recordings and common between \ac{cBSS} and \ac{SCD}. \textbf{c} Distribution of the \ac{MUAP} peak-to-peak amplitudes for the motor units commonly identified by \ac{cBSS} and \ac{SCD}, and those uniquely found by \ac{SCD}. \textbf{d} Distribution of the exponents for the intramuscular and the surface recordings.}
    \label{fig:results_real}
\end{figure}

Across the force levels, \ac{SCD} identified 41.6±12.1 motor units for subject 1, and 12.0±5.3 for subject 2, a significant increase (\(p\)\textless0.001) with respect to \ac{cBSS} (13.7±3.1 and 2.1±0.7 respectively). Additionally, the number of matched motor units between HD-i\ac{EMG} and \ac{HD-sEMG} was greater for \ac{SCD} than for \ac{cBSS} (Figure \ref{fig:results_real}, \textbf{a}). This was likely due to the fact that \ac{SCD} could identify deeper or smaller sources that were not separable with \ac{cBSS} but that were captured by the intramuscular multi-electrode arrays.

To assess decomposition accuracy of the two compared methods, the \ac{RoA} for the motor units found by both \ac{cBSS} and \ac{SCD}, and matched with HD-i\ac{EMG}, was calculated. The \ac{RoA} for \ac{SCD} was 99.1\% (\ac{IQR} 96.5\% - 100.0\%), compared to 94.6\% (\ac{IQR} 92.6\% - 98.9\%) for \ac{cBSS}. 
Figure \ref{fig:results_real} panel \textbf{b} shows the distribution of the \ac{RoA} for the commonly identified motor units, which shows a shift toward higher values for \ac{SCD} (although there was only a trend for this shift; \(p\)=0.08). Additionally, when comparing the motor units identified by both \ac{cBSS} and \ac{SCD}, as well as those uniquely identified by \ac{SCD}, it was observed that \ac{SCD} successfully detected motor units with significantly lower peak-to-peak \ac{MUAP} amplitudes (\(p\)\textless0.001), underscoring its ability to identify smaller motor unit activity that may have been overlooked by \ac{cBSS}.

Additionally, the exponent for \ac{HD-sEMG} was significantly higher (\(p\)\textless0.001) than that for HD-i\ac{EMG} (Figure \ref{fig:results_real}, \textbf{d}). 

\subsection{Experiments: Sex differences in motor unit yield (TA)}
\begin{figure}[!ht]
    \centering
    \includegraphics[ width=\columnwidth]{ 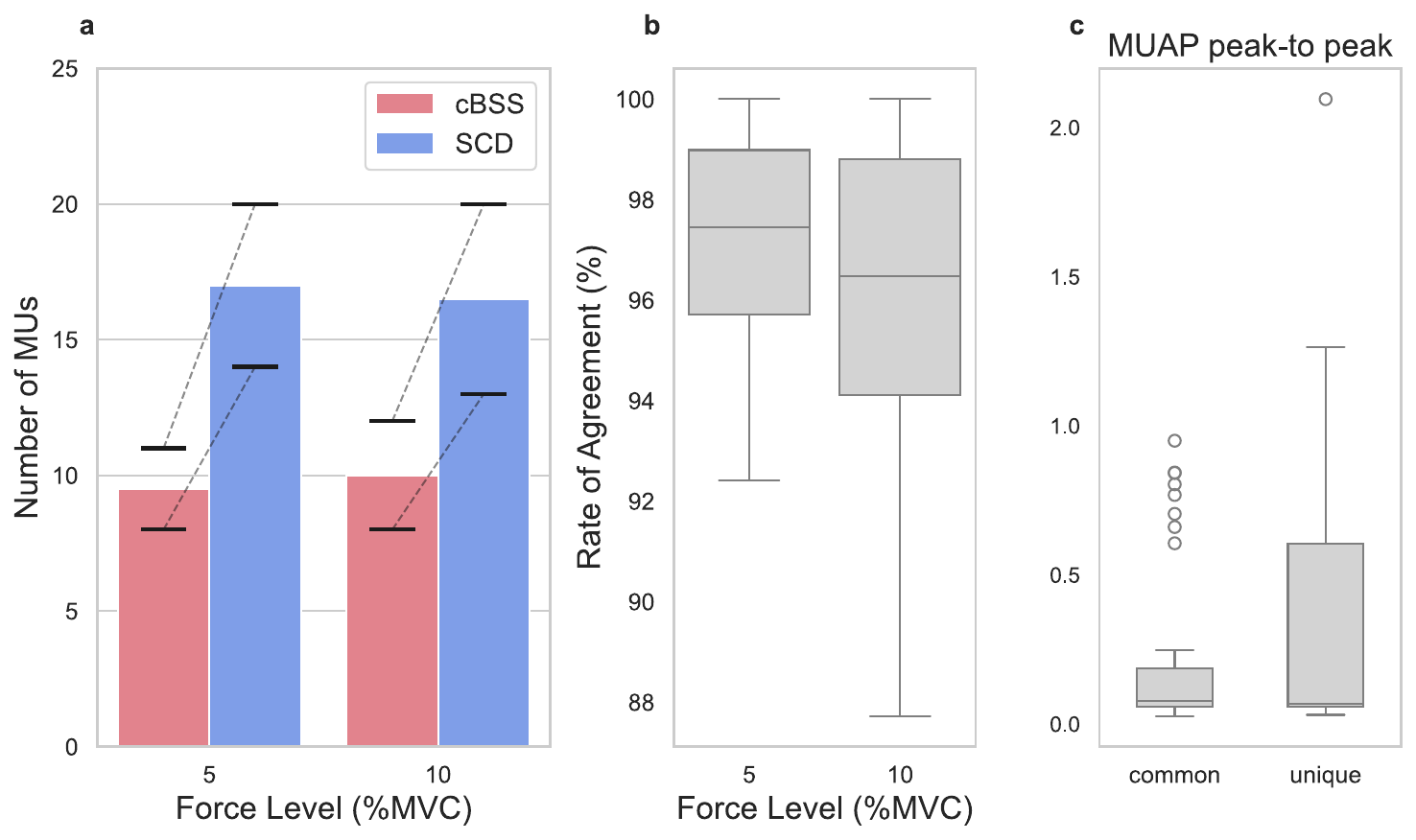}
    \caption{Effect of the decomposition algorithm used on experimental data. \textbf{a} Number of motor units found against the force level for \ac{SCD} and \ac{cBSS}. The black horizontal lines represent the value for each subject. The dashed lines connect the subjects across force levels. \textbf{b} Distribution of the \ac{RoA} for the motor units that matched between \ac{cBSS} and \ac{SCD}. \textbf{c} Distribution of the \ac{MUAP} peak-to-peak amplitudes for the motor units commonly identified by \ac{cBSS} and \ac{SCD}, and those uniquely found by \ac{SCD}.}
    \label{fig:results_females}
\end{figure}

In this experiment, the performance of \ac{SCD} and \ac{cBSS} were compared when decomposing signals recorded from female subjects. Figure \ref{fig:results_females} shows the number of decomposed motor units per subject stratified by force level (panel \textbf{a}) and the \ac{RoA} between the motor units identified by the two methods (panel \textbf{b}). On average across subjects and force levels, \ac{SCD} identified 16.8±3.8 motor units (20 and 14 for 5 \%MVC, 20 and 13 at 10 \%MVC for each subject respectively), while \ac{cBSS} identified 10.0±1.8 motor units (11 and 8 at 5 \%MVC, and 12 and 8 at 10 \%MVC for each subject respectively). Importantly, \ac{SCD} identified all the motor units identified by \ac{cBSS}, and additional ones. The \ac{RoA} between \ac{SCD} and \ac{cBSS} was 97.5\% (\ac{IQR} 95.7\% - 98.9\%) for 5 \%\ac{MVC}, and 95.7\% (\ac{IQR} 93.8\% - 98.8\%) for 10 \%\ac{MVC} (Figure \ref{fig:results_females}, \textbf{b}). The differences in the peak-to-peak amplitudes of the \ac{MUAP}s between the units commonly found by both \ac{cBSS} and \ac{SCD} and those uniquely identified by \ac{SCD} were not significant (panel \textbf{c}).

\subsection{Experiments: Motor unit yield in complex muscle groups (forearm)}
\begin{figure}[!ht]
    \centering
    \includegraphics[ width=\columnwidth]{ 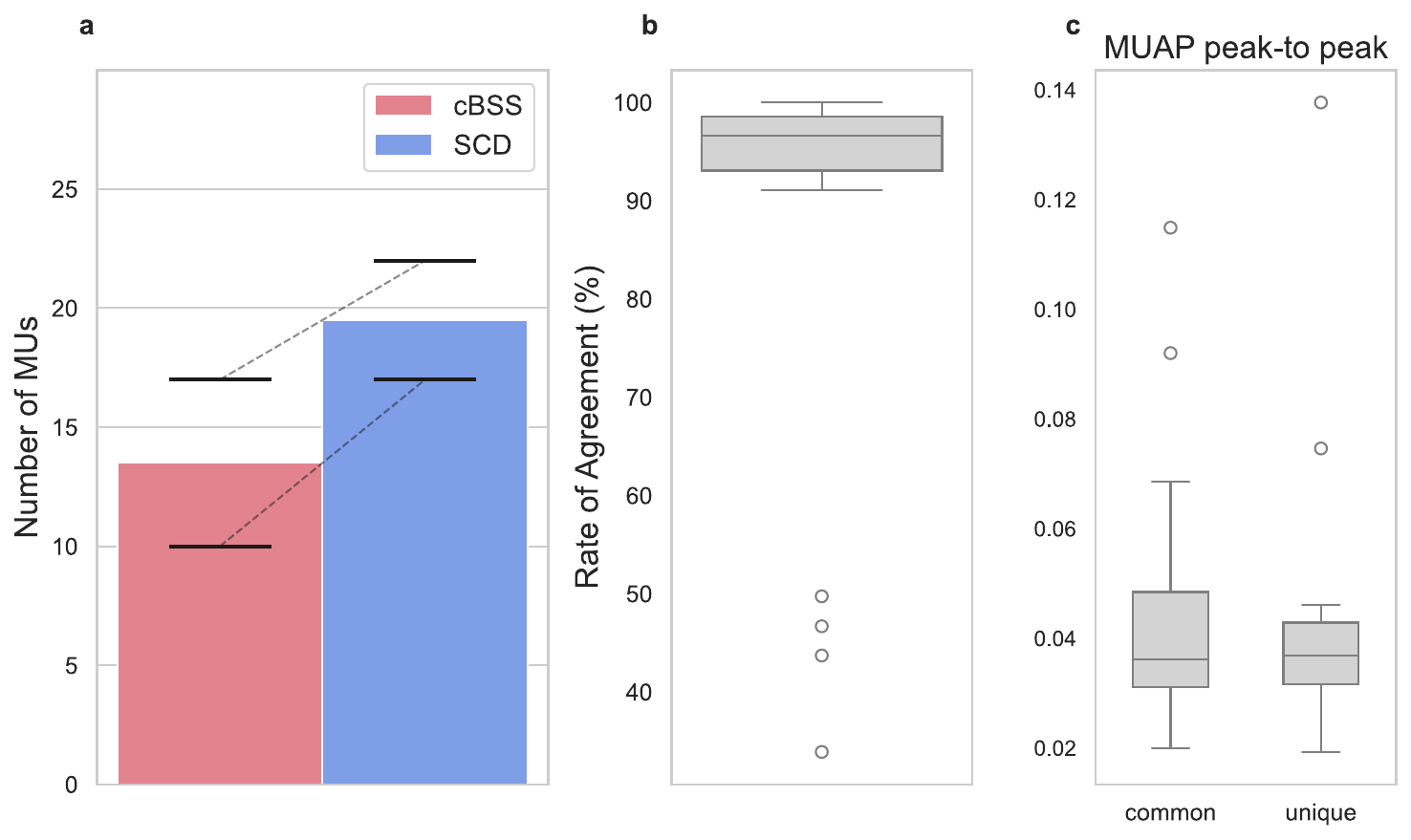}
    \caption{Effect of the decomposition algorithm used on experimental data recorded at the forearm of two female participants. \textbf{a} Number of motor units found for \ac{SCD} and \ac{cBSS}. The black horizontal lines represent the value for each subject. The dashed lines connect the subjects across force levels. \textbf{b} Distribution of the \ac{RoA} for the motor units that matched between \ac{cBSS} and \ac{SCD}. \textbf{c} Distribution of the \ac{MUAP} peak-to-peak amplitudes for the motor units commonly identified by \ac{cBSS} and \ac{SCD}, and those uniquely found by \ac{SCD}.}
    \label{fig:results_forearm_real}
\end{figure}

In this experiment, the performance of \ac{SCD} and \ac{cBSS} were compared when decomposing signals recorded from the forearm muscles of female subjects. Figure \ref{fig:results_forearm_real} shows the number of decomposed motor units (panel \textbf{a}) and the \ac{RoA} between the motor units identified by the two methods (panel \textbf{b}). For each subject, \ac{SCD} identified 17 and 22 motor units, while \ac{cBSS} identified 10 and 17 motor units respectively. Only four units that were identified by \ac{cBSS} were not identified by \ac{SCD}. Across the units with high level of agreement, the \ac{RoA} between \ac{SCD} and \ac{cBSS} was 96.9\% (\ac{IQR} 95.1\% - 98.8\%) (Figure \ref{fig:results_forearm_real}, \textbf{b}). The differences in the peak-to-peak amplitudes of the \ac{MUAP}s between the units commonly found by both \ac{cBSS} and \ac{SCD} and those uniquely identified by \ac{SCD} were not significant (panel \textbf{c}).

\clearpage
\section{Discussion}
We proposed and validated \ac{SCD} for the decomposition of \ac{HD-sEMG} signals, demonstrating its superior performance over state-of-the-art approaches. As demonstrated with both simulations and experimental results, \ac{SCD} represents a major step forward in \ac{HD-sEMG} decomposition, with broad implications for the study of the neural control of movement. The core strength of \ac{SCD} lies in its dynamic adaptation of the contrast function via particle swarm optimization and the incorporation of a peel-off strategy for sequential source removal. These features enable \ac{SCD} to address critical challenges in \ac{HD-sEMG} decomposition, such as differentiating motor units with highly similar \ac{MUAP}s and identifying motor units with low-energy \ac{MUAP}s. Unlike conventional \ac{BSS} algorithms that rely on fixed contrast functions \parencite{negro2016multi, holobar2003surface}, the ability of \ac{SCD} to adjust its objective function allows for a greater flexibility and improved handling of diverse signal characteristics.

\ac{SCD} was validated using both simulated and experimental data. In simulation, \ac{SCD} decomposed a number of motor units nearly double those identified by classic \ac{cBSS}. The motor units identified by \ac{SCD} had smaller \ac{MUAP}s (Figure \ref{fig:static}, \textbf{d} and Figure \ref{fig:results_real}, \textbf{c}), and were therefore either deeper in the muscle tissue or smaller (Figure \ref{fig:static}, \textbf{g}). Accordingly, the motor units identified by \ac{SCD} had slower conduction velocities and fewer innervated fibers (Figure \ref{fig:static}, \textbf{e, f}).  

When further applied to simulated data during ballistic tasks (Figure \ref{fig:ballistic_source}, \textbf{c}), \ac{SCD} demonstrated a 3-fold increase in the number of decomposed motor units (Figure \ref{fig:results_ballistic}, \textbf{a}), while reaching higher accuracy than the state-of-the-art. Furthermore, simulations also demonstrated that \ac{SCD} identified a greater number of motor units than \ac{cBSS} at varying levels of noise, despite the difference between methods decreased with an increase in noise (Figure \ref{fig:results_noise}, \textbf{a}). 

The simulations also revealed that both the adaptive contrast function and the peel-off procedure significantly contributed to \ac{SCD}’s performance (Figure \ref{fig:ablations}). By incrementally removing the contributions of higher amplitude \ac{MUAP}s, \ac{SCD} was able to converge to smaller, less prominent \ac{MUAP}s that would otherwise remain undetected. Importantly, the peel-off approach did not compromise performance; in all analyses, the \ac{RoA} for \ac{SCD} consistently surpassed that of \ac{cBSS}, which employs a deflation method.

Overall, the simulations indicated high performance of the proposed method in terms of motor unit yield and accuracy in decomposition. We further validated \ac{SCD} experimentally using a two-source approach by comparing HD-i\ac{EMG} and \ac{HD-sEMG} recordings \parencite{mambrito1984technique, farina2014extraction}. Matching the decomposition results between these two modalities was crucial for validating the proposed algorithm. \ac{SCD} decomposed significantly more motor units than \ac{cBSS} from experimental signals, with a higher number of matched motor units extracted from the HD-i\ac{EMG} recordings (41.6±12.1 and 12.0±5.3 vs 13.7±3.1 and 2.1±0.7, \(p\)\textless0.001). Additionally, the \ac{RoA} for the motor units commonly identified by both \ac{SCD} and \ac{cBSS} in HD-i\ac{EMG} was greater for \ac{SCD} (Figure \ref{fig:results_real}, \textbf{b}), underscoring its superior accuracy. 
Interestingly, we observed that the distribution of exponents required in the contrast function had greater mean value for \ac{HD-sEMG} signals compared to HD-i\ac{EMG} signals (Figure \ref{fig:results_real}, \textbf{d}). This finding suggests the need for a higher exponent when dealing with sources that are smaller or more similar to one another, a characteristic often seen with non-invasive signals. 

Overall, the new approach significantly increased the number of detected motor units across all conditions without compromising accuracy, which actually improved. The yield varied widely depending on the conditions, with increases ranging from approximately 50\% to over 300\%, but yet an increase was achieved consistently in all cases, across a broad range of scenarios. This consistent increase implies we can now achieve better results in conditions that previously yielded borderline unit detections. 

The method is particularly useful in conditions where the number of successfully decoded units is typically small. For instance, in cases where the \ac{MUAP}s are highly similar—such as when recording from deep muscles or from muscles covered by a large layer of subcutaneous fat— when discriminating between units becomes a major challenge. In these scenarios, standard decomposition techniques may fail to distinguish \ac{MUAP}s, often leading to failure in the decomposition. By dynamically optimizing the separation of sources, \ac{SCD} is a step forward to mitigate this problem, maximizing the contrast between similar \ac{MUAP}s and ensuring a higher yield of distinct motor units even in such complex scenarios. An example of such conditions is the decoding of motor units from female individuals. The proposed method demonstrated superior motor unit yield, as highlighted earlier. This suggests that the technique is particularly effective in addressing physiological differences, enhancing the overall accuracy and applicability of motor unit detection across diverse populations. 

We presented representative results from female individuals for both the TA muscle, typically known for successful decomposition, and the forearm muscles, where decomposition tends to be less effective.  The proposed method consistently improved the yield of decomposed motor units with respect to the state-of-the-art \ac{cBSS} in these subjects for both muscle groups.
In most conditions, in female individuals, we could enhance the number of decomposed units by around 50\%. While this does not match the numbers observed in males, it represents a substantial improvement in many cases. In some instances, adding 5 to 6 reliably extracted motor units, as shown in this study, can have a significant impact on the physiological interpretations. 

Increasing the number of concurrently decoded motor units is critical for advancing our understanding of the neural control of movement. For example, a larger pool of concurrently sampled motor units provides more comprehensive information on the distribution of common synaptic inputs to motor neuron populations \parencite{farina2014effective}. This is particularly valuable for characterizing how motor neurons are grouped to generate force and for predicting the net muscle force generated by the population of active units \parencite{caillet2023motoneuron}. Indeed, the association between estimated neural drive to muscle and muscle force output becomes stronger and more reliable as more motor units are included in the cumulative analysis.

The expansion in motor unit sample size not only enhances the precision of neuromuscular research but also opens up new possibilities for clinical and assistive applications, such as refined neural control strategies in prosthetics and advanced neural interfaces for rehabilitation \parencite{farina2017man, barsakcioglu2020control, gogeascoechea2020interfacing, tanzarella2023neuromorphic}. In these applications, adapting \ac{SCD} for real-time, online decomposition is a key future direction. 

However, \ac{SCD} has been validated exclusively in scenarios where the muscle geometry remains static. Its current framework does not account for dynamic changes in muscle fiber length or significant shifts in firing patterns. Consequently, \ac{SCD}'s performance is likely to degrade in more complex conditions, such as during dynamic muscle contractions where fiber lengths or recruitment strategies change over time. This limitation highlights the need for further development to adapt the algorithm for non-stationary conditions, where changes in muscle architecture and firing behavior could substantially impact decomposition accuracy. 

\section{Conclusions}
This study establishes \ac{SCD} as the new benchmark for \ac{HD-sEMG} decomposition. By dynamically adapting its contrast function and adopting a unique peel-off strategy, \ac{SCD} consistently outperforms traditional methods, particularly in detecting small and deep motor units. These findings, validated through comprehensive simulations and experimental data, pave the way for \ac{SCD}'s use as a new tool for the study of the neural control of movement as well as for applications ranging from clinical diagnostics to advanced human-machine interfaces. Its ability to resolve finer differences between \ac{MUAP}s marks a significant leap forward in capturing the full complexity of neuromuscular activity. 

The code used in this study is available at https://github.com/AgneGris/swarm-contrastive-decomposition.

\section{Acknowledgments}

We would like to thank the participants who took part in this study. We are also grateful to Dr. Alejandro Pascual Valdunciel for providing part of the data utilized in this study. 
AG was supported in part by UK Research and Innovation [UKRI Centre for Doctoral Training in AI for Healthcare grant number EP/S023283/1] and in part by Huawei Technologies Research \& Development (UK) Limited. IMG was supported by the EPSRC Doctoral Prize Fellowship. SM was supported by HybridNeuro (HORIZON-WIDERA-2021-ACCESS-03 - 101079392). JIP. was supported by project ECHOES (ERC Starting Grant 101077693), by a Consolidación Investigadora grant (CNS2022-135366) funded by MCIN/AEI/10.13039/
501100011033 and UE’s NextGenerationEU/PRTR funds. DF was supported by NISNEM (EPSRC
EP/T020970/1).

\clearpage
\section*{References}
\printbibliography[heading=none]

\end{document}